\documentclass[twocolumn,preprintnumbers,amsmath,floatfix,prl,aps,superscriptaddress]{revtex4}

\usepackage{graphicx}
\usepackage{epsfig} 
\usepackage{dcolumn}
\usepackage{bm}
\usepackage{times}

\makeatletter

\begin{document} 


\title{Nonperturbative Scaling Theory of Free Magnetic Moment Phases in
Disordered Metals}

\author{A. Zhuravlev} 

\affiliation{Institut f\"ur Theoretische Physik, Universit\" at
Hamburg, Jungiusstra\ss e 9, 20355 Hamburg, Germany}

\affiliation{Institute of Metal Physics, Ekaterinburg 620219, Russia}

\author{I. Zharekeshev} 

\affiliation{Institut f\"ur Theoretische Physik, Universit\" at
Hamburg, Jungiusstra\ss e 9, 20355 Hamburg, Germany}

\author{E. Gorelov}

\affiliation{Institut f\"ur Theoretische Physik, Universit\" at
Hamburg, Jungiusstra\ss e 9, 20355 Hamburg, Germany}

\author{A. I. Lichtenstein}

\affiliation{Institut f\"ur Theoretische Physik, Universit\" at
Hamburg, Jungiusstra\ss e 9, 20355 Hamburg, Germany}

\author{E. R. Mucciolo} 

\affiliation{Department of Physics, University of Central Florida,
Orlando, FL 32816-2385, USA}

\author{S. Kettemann} 

\affiliation{Institut f\"ur Theoretische Physik, Universit\" at
Hamburg, Jungiusstra\ss e 9, 20355 Hamburg, Germany}

\date{\today}

\begin{abstract}
The crossover between a free magnetic moment phase and a Kondo phase  in  low dimensional disordered metals with dilute magnetic impurities is studied.  
  We perform a finite size scaling analysis  of the distribution of the Kondo temperature  as obtained from a numerical renormalization group calculation of the local magnetic susceptibility and 
    from     the  
solution of the self-consistent Nagaoka-Suhl equation.
 We find a  sizable
fraction of free (unscreened) magnetic moments  
when the exchange coupling falls below a disorder-dependent critical
value $J_{\rm c}$. 
Our numerical results  show that between the free
moment phase due to  Anderson localization and the Kondo screened
phase there is a phase  where free moments occur
due to the appearance of random local pseudogaps 
at the Fermi energy whose width and power scale with the 
 elastic scattering rate $1/\tau$.
\end{abstract}



\maketitle


The Kondo problem is of central importance for understanding
low-temperature anomalies in low-dimensional disordered metals
\cite{dcmreview, miranda, bhatt, zu96, jetpletters, micklitz05,
meraikh, lw04, kbu042, kbu03, grempel, micklitz06}, such as
the saturation of the dephasing rate \cite{bergmann} and the
non-Fermi-liquid behavior of certain magnetic alloys
\cite{bernal,dcmreview}. For a clean metal, the screening of a
spin-$1/2$ magnetic impurity is governed by a single energy scale, the
Kondo temperature $T_K$. Thermodynamic observables
and transport properties 
obey universal functions which scale with $T_K$. Thus, in a metal
where nonmagnetic disorder is also present, two fundamental questions
naturally arise: (i) Is the Kondo temperature modified by nonmagnetic
disorder? (ii) Is the one-parameter scaling behavior still valid? It
is well known \cite{miranda,meraikh,dcmreview,jetpletters,grempel}
that magnetic moments remain unscreened when conduction electrons
are localized due to disorder. However, in weakly disordered
two-dimensional electron systems, the localization length is
macroscopically large, and so is the number $N_{\rm c}$ of 
eigenstates with a finite amplitude at the position of the magnetic
impurity. In this case, one does not expect to find unscreened
magnetic moments for experimentally relevant values of the exchange
coupling $J$.

Another situation where magnetic moments remain free in metals at low
temperatures occurs when the density of states has a {\it global
pseudogap} at the Fermi energy $E_{\rm F}$, namely, $\nu(E)\sim
(E-E_{\rm F})^{\alpha}$, where $\alpha>0$ \cite{gapless}. In clean
metals, the pseudogap quenches the Kondo screening when $J$ falls
below a critical value $J_{\rm c}(\alpha)$. So far, only a few values
of $\alpha$ have been realized experimentally: $\alpha=1$ in graphene
and in $d$-wave superconductors and $\alpha=2$ in $p$-wave
superconductors. In this Letter we examine the quantum phase diagram
of magnetic moments diluted in two-dimensional disordered metals using
a modified version of the numerical renormalization group (NRG)
method. 
We find a free moment phase which we attribute to the random occurrence of
{\it local pseudogaps}. The existence of free moments is confirmed directly with
NRG by the Curie-like behavior of the the local magnetic susceptibility
at low temperatures. Finite-size scaling is performed to
demonstrate the robustness of our finding. Furthermore, 
the distribution of Kondo temperatures  obtained numerically from NRG is found to agree well with earlier results based on the  
 solution of the Nagaoka-Suhl equation \cite{jetpletters,grempel}.

We consider the Kondo Hamiltonian of a magnetic impurity diluted in a disordered metal, 
\begin{eqnarray}
\label{hsd}
H_{\rm sd} & = & \sum_{n \sigma } E_{n}c_{n \sigma}^{\dagger }c_{n \sigma}
+ \frac{1}{2} \sum_{n,n^\prime} J_{nn^\prime} \left[ S^+ c_{n
\downarrow}^{\dagger } c_{n^\prime \uparrow} \right. \nonumber \\ & &\
\left. + S^-c_{n \uparrow}^\dagger c_{n^\prime\downarrow} +
S_z\left(c_{n \uparrow}^\dagger c_{n^\prime \uparrow} - c_{n
\downarrow}^{\dagger} c_{n^\prime \downarrow}\right) \right].
\end{eqnarray} 
Here, $E_n$ are the   eigenenergies 
of the noninteracting
conduction electrons in the disordered metal, 
which we describe by the Anderson    tight binding model,  
\begin{equation}
\label{eq:Htb}
H_{\rm V} = -t \sum_{\langle ij \rangle ~\sigma} \left( c_{i \sigma}^\dagger c_{j \sigma} + {\rm
H.c.} \right) + \sum_{i=1~ \sigma}^N V_i\, c_{i \sigma}^\dagger c_{i \sigma},
\end{equation}
with band
width $D$, nearest-neighbor hopping amplitude $t$, and random site
potentials $V_i$,
 drawn from a flat box distribution of width $W$
centered at zero. 
We consider square lattices of length
$L$ with $N=L^2$ sites and periodic boundary conditions. The exchange coupling matrix elements are then given by $J_{nn^\prime} = J\,
\psi_n^*({\bf r}) \psi_{n^\prime}({\bf r})$, with 
$\psi_{n}({\bf r})$ being the
amplitude of the single-electron eigenfunction at the position
${\bf r}$ of the magnetic impurity. Energies are
given in units of $t$.

\begin{figure}[t]
\includegraphics[width=5.cm]{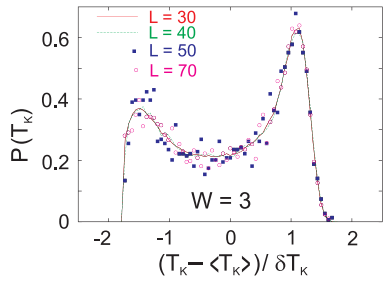}
\caption{(Color online) The distribution function of Kondo
temperatures obtained from the NSE [Eq. (\ref{eq:FTK})] for electrons
in tight-binding square lattices with disorder $W=3$. Ensembles of
$100$ samples of size $L=30$, 5 samples for $40$, and one sample for
$L= 50,70$ were used. The exchange coupling was fixed to $J/D=0.35$.}
\label{fig:kondodistribution}
\end{figure}

{\it Finite-Size Scaling.} First, we obtain all eigenenergies and
eigenfunctions of $H_V$ by using state-of-the-art numerical
diagonalization techniques for a large set of realizations of the
potential $V$ \cite{zharekeshev}. Within one-loop approximation, the
Kondo temperature is then obtained from the solution of the
Nagaoka-Suhl equation (NSE) \cite{suhl},
\begin{equation}
\label{eq:FTK}
1= \frac{J}{2 N} \sum_{n=1}^N \frac{L^2 |\psi_n ({\bf r})
|^2}{E_n-E_{\rm F}} \tanh \left(  
\frac{E_n-E_{\rm F}}{2 T_K} \right). 
\end{equation}
 We assume that $E_{\rm F}$ is between energy
levels, with even  number of electrons in the ground state of the
system. For odd number of electrons, the single electron at
the Fermi energy forms a singlet with the magnetic impurity with
binding energy $J$. Solving Eq. (\ref{eq:FTK}) numerically, one finds
strong deviations from the Gaussian behavior even for weak disorder
\cite{jetpletters}. Figure \ref{fig:kondodistribution} shows for $W=3$
that a double-peak structure persists as the lattice size is increased
from 900 to 4,900 sites. This structure
evolves into a power-law divergence in the strong-disorder limit \cite{jetpletters,grempel}. In order to verify
that these features are not an artifact of the one-loop approximation,
we performed a comparative analysis of the Kondo temperatures
obtained with the nonperturbative NRG method for different system
sizes and disorder strengths.

\begin{figure}
\includegraphics[width=8.6cm]{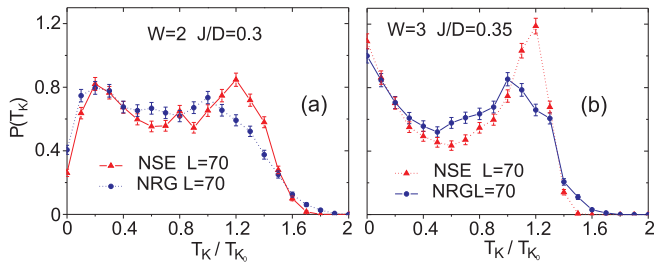}
\caption{(Color online) The distribution of Kondo temperatures
obtained with NRG for  lattices of size $L=70$. (a) $W=2$,
 $J/D=0.3$. (b) $W=3$, $J/D=0.35$ for one realization of the
disorder potential. For comparison we plot the distribution obtained
from NSE. Higher loop corrections are accounted
for by rescaling $T_K$ with $0.7 \sqrt{J/D}$. The error bars denote
the statistical error.}
\label{fig:nrgnagaokaptk}
\end{figure}

\begin{figure}[t]
\includegraphics[width=8cm]{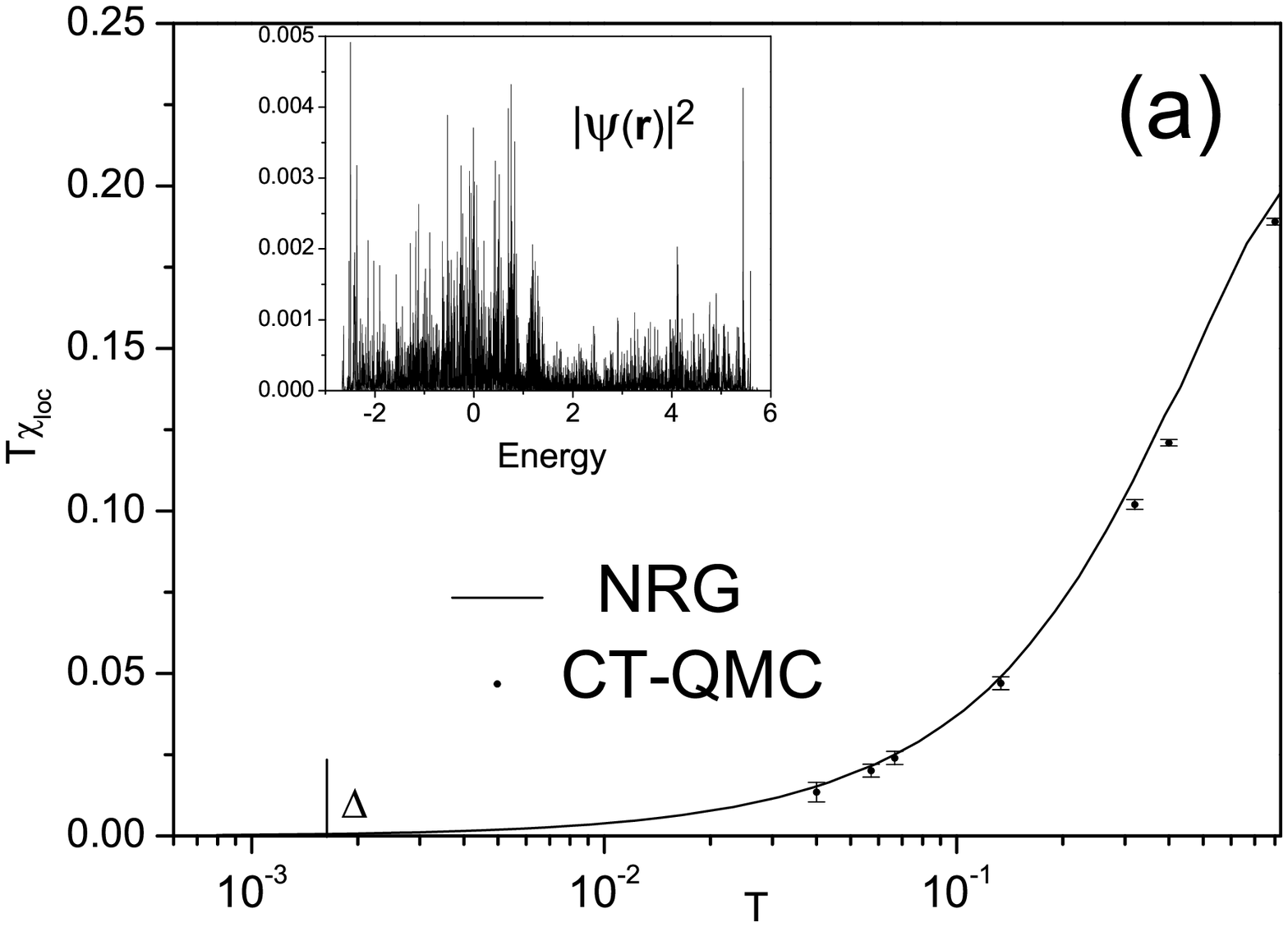}
\includegraphics[width=8cm]{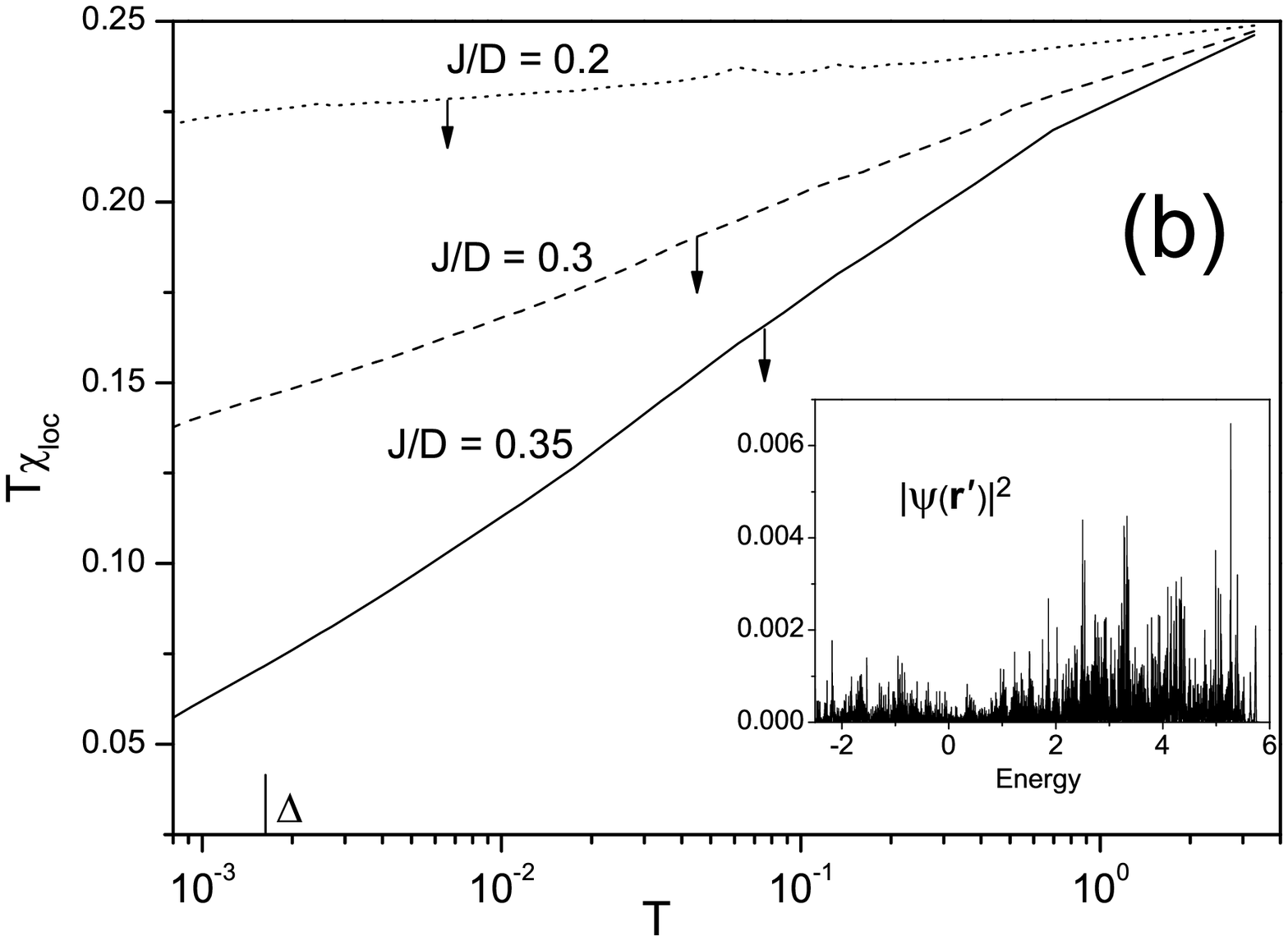}
\caption{The local spin susceptibility as function of temperature
calculated with NRG for exchange coupling $J/D=0.35$, disorder
amplitude $W=2$, and lattice size $L=70$. (a) At a site where $T_K$ is
maximal. Dots: Results obtained with the continuous-time Quantum Monte
Carlo method (CT-QMC) \cite{rubtsov,gorelov}. (b) At a site where the
magnetic impurity remains free. The arrows indicate the Kondo temperature of the clean
system for the particular value of $J/D$. Insets: The absolute value
square of the eigenfunction amplitudes at those sites as function of
energy $E$, where $E=0$ denotes the Fermi energy. The mark indicates
the mean level spacing $\Delta$. Energies and temperatures are given
in units of $t$.}
\label{fig:impsusc2}
\end{figure}

{\it NRG for Disordered Systems}. In order to apply the NRG method to
disordered systems, we use as a basis set the eigenfunctions and
eigenenergies of the the tight-binding model, Eq. (\ref{eq:Htb}), as
obtained by the numerical exact diagonalization. We perform the Wilson
logarithmic discretization of the conduction band and replace the
interval of eigenenergies between $(D/2)\Lambda ^{-n+1}$ and
$(D/2)\Lambda ^{-n},\ n=0,1,2,...$, by one energy level equal to the
average of eigenenergies in this interval. The NRG discretization
parameter is set to the smallest tractable value, $\Lambda
=1.5$. Next, with the help of the well-known tridiagonalization
procedure, we numerically map this model onto a discrete Wilson
chain. Then, we use an iterative diagonalization procedure, keeping
1,500 states after each iteration \cite{wilson}. We calculate the
local magnetic susceptibility $\chi_{\rm loc}(T)$, which is
proportional to the local correlation function of the impurity
spin. The susceptibility is found to have a smooth temperature
dependence (see Fig. \ref{fig:impsusc2}) which can be used to define
$T_K$ through the following criterion: $T \chi_{\rm loc}(T)$ crosses
over from the free spin value $1/4$ to a decay linear in $T$ as the
temperature is lowered below the Kondo temperature $T_K$. For a clean
flat band system, $\chi_{\rm loc}(T)$ coincides with the impurity
susceptibility $\chi_{\rm imp}$, which is obtained as the difference
between the susceptibility of the electronic system with and without
the impurity \cite{clogston,hewson}. Wilson defined the Kondo temperature as
the crossover temperature where $T_K \chi_{\rm imp}(T_K)=0.07$
\cite{wilson,krishna}. For disordered systems, we find that $T
\chi_{\rm imp}(T)$ can strongly deviate from the universal scaling
curve of the clean system \cite{santoro,kbu042}. It turns out,
however, that this is an artifact of the definition of $\chi_{\rm
imp}$: The susceptibility of the conduction  electrons fluctuates widely and
can result in negative values of $\chi_{\rm imp}$ \cite{gorelov}.
Therefore, we consider the local magnetic susceptibility as obtained by
differentiating the magnetic moment of the impurity with respect to a
magnetic field that acts only at the impurity \cite{ingersent}. In
Fig. \ref{fig:nrgnagaokaptk}, we show the distribution of Kondo
temperatures when $T_K$ is obtained from the Wilson condition $T_K \chi_{\rm
loc}(T_K)=0.07$. The data was extracted from a single sample of size
$L=70$ using two distinct sets of disorder and exchange coupling
amplitudes: $W=2$ and $J=0.3\, D$ (Fig. \ref{fig:nrgnagaokaptk}a), and
$W=3$ and $J=0.35\, D$ (Fig. \ref{fig:nrgnagaokaptk}b). For
comparison, we also plot the distribution of $T_K$ obtained from the
solution of the one-loop NSE, where we accounted for the known
higher-loop correction by rescaling $T_K$ with $0.7 \sqrt{J/D}$. The
agreement is remarkable and demonstrates that the double peak
structure is not an artifact of the one-loop approximation.
In Fig. \ref{fig:impsusc2}a, we plot the local impurity spin
susceptibility, multiplied by temperature $T$,  for the site with maximal $T_{\rm K}$ for a given
realization of the disorder.  In order to check that the modified   
 version of  NRG is well suited for disordered systems, 
  we  also plot in Fig. \ref{fig:impsusc2}a   
   results obtained with the continous time  quantum Monte Carlo (CT-QMC) 
    method (dots) ~\cite{rubtsov}.  For temperatures close to the 
     Kondo temperature  both methods agree well.
    The  small deviations seen  at larger temperatures can be attributed to  the fact that 
 the           average of the wavefunction amplitudes in each 
           Wilson discretization interval  results in   stronger suppression     of fluctuations at 
           higher energies.
     We note that the local density of states
(LDOS), which is shown in the inset, has most of its weight in the
lower half of the band, close to the Fermi energy. In contrast, in
Fig. \ref{fig:impsusc2}b we show the impurity susceptibility times $T$ for a
site where the magnetic impurity remains unscreened for
$J/D<0.35$. Note that $T \chi_{\rm loc}$ changes only weakly with
temperature in this case. For $T \ll \Delta$ it approaches its free
value $1/4$ (not shown). The inset shows the corresponding LDOS where
one sees that not only the weight is shifted towards the upper half of
the band, away from the Fermi energy (which is set at quarter
filling), but also that there is a minimum in the LDOS  at the Fermi energy,
resembling a {\it local pseudogap}.

\begin{figure}
\includegraphics[width=8.5cm]{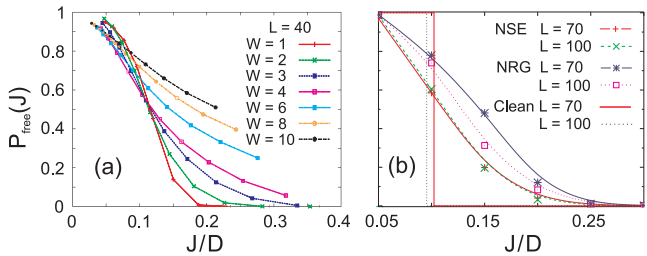}  
\caption{(Color online) Fraction of free moments as function of 
exchange coupling obtained by two  methods: (a) From the
condition that the NSE has no solution; (b) from  temperature
dependence of local spin susceptibility calculated with  NRG for
disorder $W=2$. The corresponding NSE result, and the fraction of free
moments in a clean sample (step functions) are plotted for
comparison. Lattice sizes are indicated. The lines
are guides to the eye. }
\label{fig:freemoments}
\end{figure}

\begin{figure}
\includegraphics[width=8.5cm]{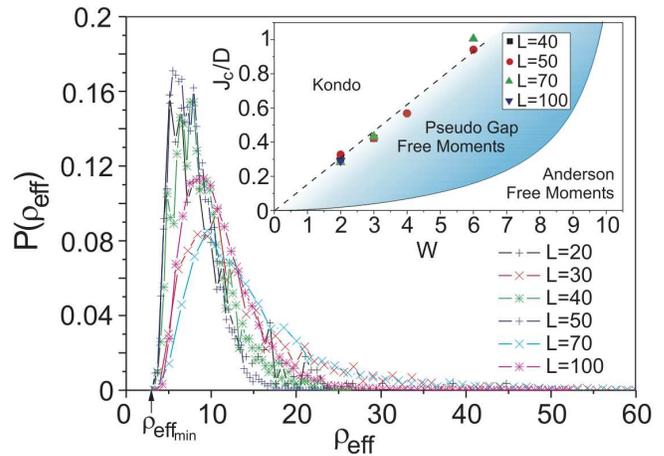}
\caption{(Color online) The distribution of $\rho_{\rm eff}$ for $W=2$
and various lattice sizes. Inset: The magnetic impurity phase diagram
in the $(J,W)$ plane. The data points correspond to $J_{\rm
c}=1/\rho_{\rm eff_{min}}$ as function of $W$, obtained
numerically for several system sizes. The straight line is a guide to
the eye. The critical exchange coupling expected  due to the hard gap
caused by Anderson localization, $J_{\rm c}^{\rm A}$ [see
Eq. (\ref{anderson})] is plotted for comparison (full line). For $J > J_{\rm c}$,
all magnetic moments are screened (Kondo phase). For $J< J_{\rm
c}^{\rm A}$, all magnetic moments remain free due to Anderson
localization, the Anderson free moment phase. In the intermediate phase  there is  a finite
probability that some magnetic moments remain unscreened due to the random occurence of local pseudo gaps.}
\label{fig:rhoeff}
\end{figure}

{\it Free Moment Phase}. Next, we use NRG to analyze in detail the impurity sites where the
renormalization of the magnetic susceptibility remains so small that
no finite Kondo temperature can be defined. On such sites a magnetic
moment remains free down to the lowest temperatures, 
the spin susceptibility $\chi_{\rm loc}$ diverges and $T \chi_{\rm
loc}(T)$ approaches a finite value, as seen in
Fig. \ref{fig:impsusc2}b. We employed this criterion to identify free
magnetic moments. In Fig. \ref{fig:freemoments}, we compare the
resulting fraction of free moments with that derived from the
condition that Eq. (\ref{eq:FTK}) has no finite $T_K$ solution. We see
that the fraction of free moments obtained from  NSE is smaller than
the one from NRG. This can be attributed to the fact that higher-loop
corrections tend to lower $T_K$. Thus, the criterion that the NSE has
no solution gives a lower bound for the fraction of free magnetic
moments. This criterion corresponds to setting $T_K
\rightarrow 0$ in Eq. (\ref{eq:FTK}). Therefore, free moments exist
when the effective, weighted LDOS as defined by,
\begin{equation} 
\rho_{\rm eff} = \frac{1}{ 2 N} \sum_{n=1}^N \frac{L^2 |\psi_n ({\bf r})
|^2}{\mid E_n-E_{\rm F}
 \mid},
\label{rhoeff}
\end{equation}
is equal or smaller than the inverse exchange coupling $1/J$. For a
fixed realization of the disorder, this yields a lower bound for the
critical exchange coupling $J_{\rm c}$ below which the magnetic
impurity remains unscreened.
For a clean system with a flat band, $\rho_{\rm eff}$ diverges
logarithmically with the number of states $N$ or, equivalently, the
volume. Then, $J_{\rm c} /D\sim 1/ \ln N$. Since $T_K \sim D
\exp(-D/J)$, the condition for free moments becomes $\Delta > c\,
T_K$, with a constant $ c \gg 1$ which depends on $E_{\rm F}$. In a
clean system the level spacing vanishes in the thermodynamic limit,
therefore there are no free moments for any finite $J$. Can one
conclude the same when nonmagnetic disorder is present? 
 No, for the following two reasons. First, let us
note that for disordered systems with dimensions $d \le 2$, all energy
eigenstates are localized in the absence of spin-orbit interaction or
a strong magnetic field. A finite local gap of order $\Delta_{\rm c}
=1/(\nu \xi^d)$ develops at the Fermi energy, where $\nu$ is the
average density of states and $\xi$ is the average localization
length. Therefore, there are with certainty free moments whenever
$\Delta_{\rm c} \gg T_K$, or, equivalently, $J \ll J_{\rm c}^{\rm A}
\sim D/\ln N_{\rm c}$, where $N_{\rm c}= D/\Delta_{\rm c}$. However,
in weakly disordered two-dimensional electronic systems $\xi (g) = g
\exp(\pi g)$ in the absence of an applied magnetic field. Here,
 the dimensionless disorder parameter $g= E_{\rm F} \tau$ is  for
quarter filling related to the disorder  amplitude $W$ as  $g=96/(\pi W^2)$. Using Eq. (\ref{rhoeff}) we then
find
\begin{equation}
\label{anderson}
J_{\rm c}^{\rm A} = D\, \left[ \ln \left( 2 \xi(W)^2 \right) + C
\right]^{-1},
\end{equation}
where $C\approx 0.58$. This expression only provides a lower bound for
$J_{\rm c}$ because both the localization length and the LDOS are
widely distributed in a disordered system. In Fig. \ref{fig:rhoeff} we
show the distribution of $\rho_{\rm eff}$ for moderate disorder and
various lattice sizes. Remarkably, the point where the distribution
drops sharply to zero, $\rho_{\rm eff_{min}}$, hardly depends on
lattice size. In the inset of Fig. \ref{fig:rhoeff} we plot $J_{c}
=1/\rho_{\rm eff_{min}}$ as a function of disorder for various system
sizes, together with $J_{\rm c}^{\rm A}(W)$,
Eq. (\ref{anderson}). This plot shows a quantum phase diagram in the
$(J,W)$ plane with a free magnetic moment phase due to Anderson localization for  $J< J_{\rm c}^{\rm A}$  and
a Kondo screened phase for $J> J_{c}(W) $. 
For intermediate couplings,  $J_{\rm c}^{\rm A} < J < J_{c}(W) $,
we also find free magnetic moments for all system sizes considered. 
 It is well known that a free moment phase exists when there
  is a pseudogap    around  the Fermi energy \cite{gapless}.  Indeed, as seen in 
   Fig. \ref{fig:impsusc2}b, there is a dip in the LDOS  at sites where the 
    magnetic moment remains unscreened. 
       In weakly disordered metals, the LDOS is correlated over
        a macroscopic energy interval of order of the elastic
         scattering rate $1/\tau$~ \cite{mirlin}. Although these correlations
          are  only of order $1/g$ and the  LDOS can fluctuate
in energy, they can still cause dips in the LDOS within a range of
 order       $1/\tau$.     
 From  the numerical results for 
   $J_c(W)$ (inset of Fig. 5)  we infer  that the  local pseudogaps
   in weakly disordered metals   
    have  a power that increases with 
     disorder as $\alpha \sim W/D \sim 1/\sqrt{g}$.


The authors gratefully acknowledge useful discussions with 
Christopher B\" auerle, Harold
Baranger, Ribhu Kaul, Ganpathy Murthy, Achim Rosch, Laurent Saminadayar, Denis Ullmo, and Gergely
Z\'arand. S.K. and E.R.M. gratefully acknowledge the hospitalities of
the Condensed Matter Theory Group at Boston University, the Max-Planck
Institute for Physics of Complex Systems, and the Aspen Center for
Physics. This research was supported by the German Research Council under SFB 668, B2, SFB 508, B9, and  KE 807/2-1.  A.Z. acknowledges support by
the Grant No. 4640.2006.2 from the Russian Basic Research Foundation and by the Ural Division of RAS.



\end{document}